\documentclass[useAMS,usenatbib]{mn2e}

\newcommand{\T}{TrES-1}
\newcommand{\hd}{HD 209458b}
\usepackage{graphics}
\usepackage{wrapfig}

\newcommand{\apj}{ApJ}
\newcommand{\apjl}{ApJ}
\newcommand{\planss}{Planet. Space Sci.}
\newcommand{\jgr}{JGR}
\newcommand{\pasj}{PASJ}
\newcommand{\aap}{A\&A}
\newcommand{\nat}{Nature}

\title[Condensates and Transmission Spectroscopy]{The Effect of Condensates on the Characterization of Transiting Planet Atmospheres with Transmission Spectroscopy}
\author[J.J. Fortney]{Jonathan J. Fortney\thanks{E-mail: jfortney@arc.nasa.gov}\\
Space Science and Astrobiology Division, NASA Ames Research Center, MS 245-3, Moffett Field, CA 94035}

\begin{document}

\date{Accepted September 7, 2005}

\pagerange{\pageref{firstpage}--\pageref{lastpage}} \pubyear{2005}

\maketitle

\label{firstpage}

\begin{abstract}

Through a simple physical argument we show that the slant optical depth through the atmosphere of a `hot Jupiter' planet is $\sim$~35-90 times greater than the normal optical depth.  This not unexpected result has direct consequences for the method of transmission spectroscopy for characterizing the atmospheres of transiting giant planets.  The atmospheres of these planets likely contain minor condensates and hazes which at normal viewing geometry have negligible optical depth, but at slant viewing geometry have appreciable optical depth that can obscure absorption features of gaseous atmospheric species.  We identify several possible condensates.  We predict that this is a general masking mechanism for all planets, not just for \hd, and will lead to weaker than expected or undetected absorption features.  Constraints on an atmosphere from transmission spectroscopy are not the same as constraints on an atmosphere at normal viewing geometry.
\end{abstract}


\section{Introduction}
To date, 8 extrasolar giant planets (EGPs) are known to transit their parent stars in tight, several day orbits.  Characterizing the atmospheres of two of these planets (\hd~and \T) has been a major goal for many astronomers in the past few years.  Not long after the discovery of the transits of planet \hd~\citep{Charb00,Henry00}, a number of studies appeared in the literature on radiative transfer aspects of `Pegasi planet' (or `hot Jupiter') transits, namely how stellar light passing through a planet's atmosphere can be absorbed at wavelengths where opacity is high.  This would lead a distant observer who obtained a spectrum of the star during a transit of the planet to see the planet's atmospheric absorption spectrum superimposed on the star's spectrum.  The physics behind this process were laid out theoretically and modeled by \citet{SS00}, \citet{Brown01}, and \citet{Hubbard01}.  The consensus of these three studies was that for \hd, for a clear, cloudless atmosphere, that the transit depth (0.016 in relative flux) could itself vary by up to a few \% due to absorption by gaseous sodium, potassium, water, and carbon monoxide.

Shortly thereafter \citet{Charb02} used the STIS instrument aboard \emph{Hubble Space Telescope} to observe the predicted sodium absorption doublet at 589 nm.  However, the magnitude of this absorption was 2-3 times weaker than had been predicted by theoretical models that assumed a clear and cloudless atmosphere.  A number of possible reasons for this discrepancy were put forth in \citet{Charb02}, including a global underabundance of sodium, ionization of sodium by stellar flux, sodium being tied up in condensates or molecules \citep[see also][]{Atreya03}, or high clouds obscuring the absorption of sodium \citep[see also][]{SS00}.  In addition, \citet{Barman02} found that neutral atomic sodium may be out of local thermodynamic equilibrium, leading to a deficit of sodium atoms able to absorb at 589 nm, relative to LTE calculations.  \citet{Fortney03} derived a self-consistent pressure-temperature (\emph{P--T}) profile for \hd, and found that silicate and iron clouds reside high in the planet's atmosphere, at the several mbar level, and that these opaque clouds mask the absorption of sodium enough to match the \citet{Charb02} observations.  However, as noted by the authors at the time, this conclusion is extremely sensitive to the \emph{P--T} profile calculated for the atmosphere.  Recently \citet{Iro05} have shown that a substantial day-night temperature contrast in the planet's atmosphere could lead to a sink of atomic sodium, as the atom could be tied up into the condensate Na$_2$S on the planet's night side.

In addition to the sodium observation, \citet{Deming05a} have recently attempted to observe absorption due to first overtone bands of CO at $\sim$~2.3 $\mu$m, using NIRSPEC on the Keck Telescope.  Their sensitively was high enough such that if CO was present in the abundances predicted for a clear and cloud-free atmosphere, that the CO should have been detected.  However, it was not, and the authors point to the masking effect of high clouds as the likely culprit.  Currently other searches are underway to detect the absorption of H$_2$O \citep{Harrington02} and H$_3^+$ \citep{Haywood04} in the atmosphere of \hd.  \citet{Narita05} have also reported upper limits due to absorption by Li, H, Fe, and Ca.  These studies are in addition to the exosphere absorption studies such as \citet{Vidal03,Vidal04}, which have yielded important results, but are not the subject of our study here.

To date, there has been no discussion in the extrasolar planets literature regarding how condensates less abundant than the `standard' brown dwarf condensates (such as silicates and iron) may effect transmission spectroscopy.  In the following sections we show that condensates that may have insignificant optical depth when viewed at normal geometry can have appreciable optical depths for the \emph{slant} viewing geometry relevant for transits. Searches for transmission absorption features in the atmosphere of \hd, or any other similar searches for absorption during transits of other planets in the future, will very likely find weak or nonexistent absorption features.  Constraints on atmospheric abundances derived from transmission spectroscopy will not map directly as constraints on abundances under normal viewing geometry. 

\section{Geometry of the Problem}
The studies of \citet{Hubbard01} and \citet{Fortney03} showed that the pressures probed by transit observations are sensitive function of wavelength.  Specifically for \hd, \citet{Fortney03} showed that in the spectral region from 580 to 640 nm, which spanned the \citet{Charb02} observations, the pressure where the slant optical depth reached 1 varied from a few $\mu$bar near the sodium line cores, to potentially 50 mbar at 640 nm, if the atmosphere was clear.  This pressure could reach $\sim$~a few hundred mbar at wavelengths of relatively low opacity.

At these low pressures it is reasonable, to first order, to approximate the atmosphere as having a constant temperature with height.  If a planet's atmosphere is in hydrostatic equilibrium, with a constant temperature and mean molecular weight, then the barotropic law holds, which states \citep{ChambHunt}:
\begin{equation} \label{pR}
p(r)=p(r_0)exp\left(-\frac{GMm}{kTrr_0}(r-r_0) \right)
\end{equation}
where $p$ is the pressure, $r$ is the radius of interest, $r_0$ is the reference radius, $G$ is the gravitational constant, $M$ is the mass of the planet, $m$ is the mean mass of a molecule in the atmosphere, $k$ is Boltzmann's constant, and $T$ is the temperature.  The vertical integrated column density of the atmosphere $N_V(r_0)$, above a given local density $n(r_0)$, is given by:
\begin{equation} \label{nH}
N_V(r_0)\equiv \int_{r_0}^{\infty} n(r)dr = \int_{0}^{p(r_0)} \frac{r^2}{GMm} dp \approx \frac{p(r_0)}{g(r_0)m} = n(r_0) H,
\end{equation}
where $H$ is the scale height ($H=kT/mg$) and $g$ is the planet's gravitational acceleration.  Clearly, if our reference density is $n_0$, our column density is then $n_0 H$.  With the assumption that $g$ is constant in the atmosphere \mbox{Eq.~(\ref{pR})} can be simplified to:
\begin{equation} \label{pZ}
p(z)\approx p(z_0)exp\left(-\frac{z-z_0}{H} \right)
\end{equation}
where here we have replaced the two radii, $r$ and $r_0$ with heights from a given level, $z$ and $z_0$, respectively.

\begin{figure}
\includegraphics{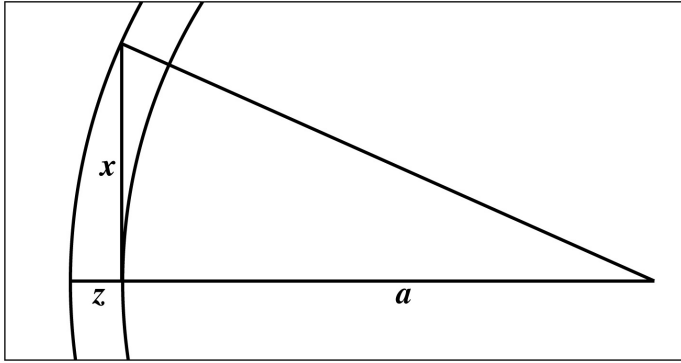}
\caption{Diagram of slant vs. normal geometry.  $a$ is the planet's radius out to a standard level, say, the radius at normal optical depth unity.  $z$ is the thickness of the atmosphere above this level, out to some very low pressure.  $x$ is the distance to this same low pressure, towards the horizon.}
\label{FIGcut}
\end{figure}

We now turn to \mbox{Fig.~\ref{FIGcut}}.  Here $a$ is a given radius of the planet, say where the normal optical depth is unity.  The thickness of our atmosphere is $z$ and $x$ is a line tangential to our optical depth unity `surface,' horizontal to the horizon.  Using the Pythagorean theorem and that $2az \gg z^2$, then $z \approx x^2/2a$.  Therefore, \mbox{Eq.~(\ref{pZ})}, when written in terms of density, rather that pressure, becomes
\begin{equation} \label{nhorz}
n(x)=n_0 exp \left(\frac{-x^2}{2aH}\right)
\end{equation}
and the horizontal integrated density $N_H$, from horizon to horizon, is
\begin{equation} \label{vertint}
N_H = \int_{-\infty}^{\infty} n(x) dx = n_0 \sqrt{2\pi a H}
\end{equation}
The ratio of the horizontal integrated density to the vertical integrated density is the quantity of interest here.  This quantity, which we will label $\eta$, simplifies to
\begin{equation}
\label{eta}
\eta = \frac{N_H}{N_V} = \frac{\tau_H}{\tau_V} = \sqrt{\frac{2\pi a}{H}}.
\end{equation}
This shows that the horizontal integrated density is significantly larger that the vertical integrated density.  This ratio is $\sim$~75 for Earth (and $\sim$~128 for Jupiter), and leads to an atmosphere that has significantly different absorption and scattering properties at a \emph{slant} geometry than for \emph{normal} geometry, which one can easily observe at sunset.  Since the optical depth is directly proportional to the column density, $\eta$ is also the ratio of the slant optical depth to the normal optical depth.
\section{Application to EGPs}
\subsection{Atmospheric Properties}
For \hd, assuming $T=1200$ K, $a=10^5$ km, and mean molecular weight $\mu=2.3$, we find $\eta = 38$.  For \T, which although similar in mass has a 30\% smaller radius and is $\sim$~300 K colder in effective temperature \citep{Fortney05}, we find $\eta = 50$.  A minor condensate, having a normal optical depth of 0.02, which would be easily ignored when calculating a planet's emission or reflection spectrum, would have an optical depth of 1 in slant transmission through the planet's atmosphere.

In \mbox{Fig.~\ref{FIGpt}} we plot self-consistent \emph{P--T} profiles for planets \hd~and \T, as taken from \citet{Fortney05}.  Also plotted are condensation curves for a variety of equilibrium condensates, spanning a large range in temperature.  The condensation curves are taken from \citet{Lodders05}.  We note that the \emph{P--T} profile shown here for \hd~is cooler than that of \citet{Fortney03} because here we assume the planet is able to reradiate absorbed stellar flux over the entire planet, whereas \citet{Fortney03} assumed this reradiation could only occur on the planet's day side, leading to a warmer day side \emph{P--T} profile.  In addition, \citet{Fortney03} utilized a different model atmosphere program \citep[see][]{Sudar03}.

We have used the transit radiative transfer program described in \citet{Hubbard01} and \citet{Fortney03} to compute both the normal optical depth and slant optical depth at various pressures in the atmosphere of both planets.  This code takes a computed model atmosphere, for which $T$, $P$, $\rho$, and extinction cross-sections are known as a function of altitude, under the assumption of hydrostatic equilibrium, and places the atmosphere onto an opaque disk of a given radius.  Normal and slant optical depths are then computed numerically through the atmosphere.  For isothermal atmospheres, we recover the same ratio $\eta$ we calculated analytically, within less than $\sim$~1\%.  For the atmospheres we consider here our pressure range of interest is the upper troposphere, and temperature decreases with altitude.  Integrating upwards in altitude from a pressure of 1 bar, we calculate $\eta=30$ for \hd~and 40 for \T, about 80\% of our simple analytic case.  However, integrating from a  lower pressure, say 10 mbar, may be more relevant, and from this pressure we calculate $\eta=35.5$ for \hd~and 47 for \T, which is $\sim$~94\% of the value obtained from our simple analytical treatment.  In summary, a more detailed analysis gives ratios of the slant optical depth to normal optical depth for these atmospheres that is consistent with our earlier simple analysis.  These values are collected in \mbox{Table 1}.

\subsection{Condensate Scale Height}
The values for $\eta$ calculated so far assume that the scale height for condensate is the same as the scale height of the surrounding gas.  However, there is strong evidence in Jupiter's atmosphere that its visible ammonia cloud is more compact than the surrounding gas.  Using Voyager infrared spectra  \citet{Carlson94} derived a ratio of the scale height of condensate ($H_{cond}$) to scale height of gas ($H$) of $0.35\pm0.10$ for Equatorial Zones and $0.40\pm0.10$ for Northern Tropical Zones.  Observations of the Jovian tropics with the \emph{Infrared Space Observatory} by \citet{Brooke98} indicate a scale height ratio of 0.3.

Further evidence for clouds of small vertical extent comes from observations of L-type stars and brown dwarfs, which have silicate and iron clouds in their visible atmospheres.  The observed spectra of these objects have been accurately modeled by \citet{Marley02} and \citet{Marley04} using the 1D cloud model of \citet{AM01} with a sedimentation efficiency parameter, $f_{sed}$= 2-3.  This $f_{sed}$ range gives silicate and iron clouds with $H_{cond}/H=0.25-0.30$ \citep{AM01}.  These observations and models indicate that equilibrium condensates, across a wide range of temperatures and chemical species, have scale heights that are $\sim$~1/3 of the local gas scale height.  This leads to values of $\eta$ that are $\sim$~75\% larger than calculated earlier.  Values of $\eta$ would then increase to 66 for \hd~and 87 for \T.

\subsection{Minor Condensates and Hazes}
\begin{figure}
\includegraphics{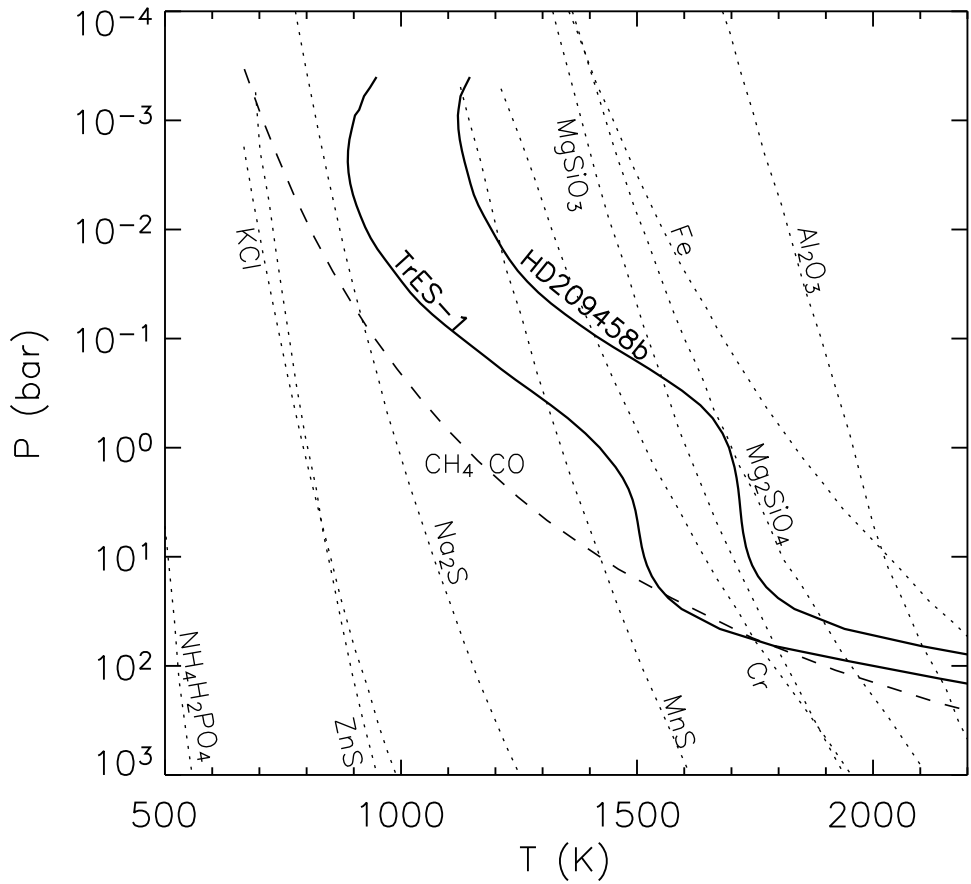}
\caption{Pressure-temperature profiles for \T~and \hd.  Condensation curves for various compounds, as taken from \citet{Lodders05} are shown as dotted lines.  The boundary where CH$_4$ and CO have the same abundance is shown as a dashed line.}
\label{FIGpt}
\end{figure}

For brown dwarf atmospheres, corundum (Al$_2$O$_3$), iron (Fe), and silicates (MgSiO$_3$ and/or Mg$_2$SiO$_4$) appear to be the only condensates that have appreciable optical depth, and therefore leave any imprint on the spectra of these objects.  \citep[See, for instance,][]{AM01, Marley02, Cooper03, Allard01, Lodders05}.  However, at slant viewing geometry, one likely has to consider condensates that may be a factor of 100-1000 less abundant, compared to the silicates.  From the work of \citep{Fegley94}, on condensation in the deep atmospheres of Jupiter and Saturn, and \citet{Lodders03}, on the condensation temperatures of the elements, we can highlight condensates that fit into this abundance range.  In order of decreasing condensation temperature, these are: Cr, MnS, Na$_2$S, ZnS, KCl, and NH$_4$H$_2$PO$_4$.  These are the condensation curves plotted in \mbox{Fig.~\ref{FIGpt}}.

To analyse the potential optical depths of these species, we will closely follow the analysis of \citet{Marley2000}, who derived an expression for the maximum optical depth of condensates in substellar atmospheres.  We will use this equation to determine the relative optical depths of various species.  This expression is:
\begin{equation}
\label{maxtau}
\tau_{\lambda} = 75\epsilon Q_{\lambda}(r_c)\varphi \left(\frac{P_c}{1\textrm{ bar}}\right) \left(\frac{10^5 \textrm{cm sec}^{-2}}{g}\right) \left(\frac{1 \mu\textrm{m}}{r_c}\right) \left(\frac{1.0\textrm{g cm}^{-3}}{\rho_c}\right),
\end{equation}
where $\epsilon$ is a factor $\le 1$ that accounts for the finite amount of species left over after condensation, because of the vapor pressure above the condenstate, $Q_{\lambda}$ is the wavelength dependent extinction efficiency from Mie theory, $r_c$ is the radius of the condensate particles, and  $\varphi=f m_c / \bar{m}$, where $f$ is the mixing ratio of the species, $m_c$ is the molecular weight of the condensed species, $\bar{m}$ is the mean molecular weight of the atmosphere.  Additionally, $P_c$ is the condensation pressure, $g$ is gravitational acceleration in the atmosphere, and $\rho_c$ is the mass density of the condensate.  Since at this point we are only interested in the \emph{relative} optical depths of the various species at a given $P_c$, we will make several simplifications.  Following Marley, we assume that $\epsilon$, $Q_{\lambda}$, $r_c$, $\rho_c$ are approximately equal for the condensate species, then the optical depth ratio for given condensates 1 and 2 reduces to
\begin{equation}
\label{ratio}
\frac{\tau_1}{\tau_2} = \frac{f_1 m_{c1}}{f_2 m_{c2}}.
\end{equation}
For illustrative purposes, we used a \emph{P--T} profile for \hd~that is slightly warmer than that shown in \mbox{Fig.~\ref{FIGpt}}, along with the \citet{AM01} cloud model with $f_{sed}=2$, to compute the normal optical depth of a MgSiO$_3$ cloud with a base at 30 mbar.  We find that the optical depth is 0.5 at normal viewing geometry at a wavelength of 1 $\mu$m.  Based on this calculated value, we then determine normal optical depths for our other condensates, if they were to form at this pressure, using \mbox{Eq.~\ref{ratio}}.  We also calculated the slant optical depths for \hd, assuming $\eta=66$.  These are listed in \mbox{Table 2}.  The $f$ values for each condensate is the $f$ of the limiting atomic species for each condensate, as given for `Solar System Abundances' in \citet{Lodders03}.  This assumes that there are not other molecules tying up the atoms, an assumption that is accurate at these high temperatures.  The slant optical depths calculated are on the order of $\sim$~0.1-1, meaning their opacity is not neglible.  From this analysis we can see that these minor condensates may well mask absorption features due to gaseous atomic and molecular species.  In addition, if the atmospheres of transiting planets are enhanced in heavy elements these optical depths could well be larger.  Jupiter's atmosphere is enhanced in heavy elements by a factor of about 3 times over solar composition \citep{Atreya03} and Saturn's methane abundance has recently been pinned at $\sim$7 times solar \citep{Flasar05}.

Non-equilibrium hazes, such as the photochemically-produced hydrocarbon hazes found in the atmospheres of Jupiter, Saturn, Uranus, Neptune, Titan, and Los Angeles, have long been observed and modeled \citep[e.g., for Jupiter:][]{West86, Tomasko86, Rages99}.  In Jupiter these stratospheric hazes can have normal optical depths of a few tenths at high latitudes.  The scale height of these hazes is generally similar to the scale height of the surrounding gas \citep{Moses95,Rages99}.

To date, only \citet{Liang04} have studied whether photochemically-produced hydrocarbon hazes will be found in Pegasi planet atmospheres.  These authors found it very unlikely that Pegasi planets would have hazes of this sort, due to several reasons.  These include:  a lack of methane to be photolized (CO is the dominant carbon carrier), the high atmospheric temperatures, which would not allow any hydrocarbon products that were formed to condense, and fast reverse reactions that quickly break down hydrogenated carbon compounds.  However, these authors acknowledge they do not consider ion-neutral chemistry, and they also note that the stellar wind could be a vast source for high-energy charged particles.  Much work still needs to be done to definitively rule out hazes not predicted by equilibrium chemistry.  Even condensates that form relatively thin hazes could have important effects on transmission spectroscopy.

\section{Discussion and Conclusions}
In our calculations we have used a relatively simple atmosphere model.  We assume that the limb of the planet is uniform around the entire planet and that the atmosphere has a uniform \emph{P--T} profile.  It is certainly possible that this condition will not be met in reality \citep[see][]{Showman02,Cho03,Cooper05,Barman05}.  As \citet{Iro05} showed in the region of the 589 nm sodium doublet, it is possible for `hot' and `cold' limbs of a planet to show significantly different transmission signatures.  Our main argument will likely hold even if the distribution of condensates is more complex than we have assumed.  For instance, if the temperature on the limb simply monotonically increases from the night to day sides, one could imagine a series of atmosphere profiles where a given condensation curve is crossed at progressively lower pressures.  At a given pressure at the terminator, this could lead to a lower condensate slant optical depth on the night side, but a higher slant optical depth on the day side, relative to our treatment here.  It is certainly possible that different condensates could be important in different locations in a planet's atmosphere.

In this paper we have taken a straightforward look at the optical depth that minor condensates may have in the slant viewing geometry relevant to planetary transits.  While our findings are potentially not unanticipated, we felt the need to discuss this issue because the impact of minor condensates on the transit characterization of planetary atmospheres had not been discussed in the extrasolar planet literature to date.  Our conclusions can be summarized as follows:
\begin{itemize}
\item For the \emph{slant} viewing geometry relevant to transmission spectroscopy observations of EGPs, the slant optical depth can be on the order of 35-90 times larger than the normal optical depth.  This depends upon the scale height of the condensate specifically, which may be smaller than the scale height of the surrounding gas.
\item Constraints on cloud location and thickness, and/or constraints on chemical abundances, obtained from transmission spectroscopy will not map directly onto constraints for the atmosphere when viewed at normal geometry.  A cloud can be optically thick at slant viewing geometry and optically thin at normal viewing geomery.  Thus, there could be abundant atomic Na and CO in the atmosphere of \hd, even though \citet{Charb02} observed only a weak Na absorption feature, and \citet{Deming05a} observed no CO.  The obscuring opacity source for this planet could be condensed Cr, MnS, silicates, or Fe, and will depend on the actual temperatures on the planet's limb.
\item Minor equilibrium condensates or photochemically derived hazes that may be reasonably ignored for normal viewing geometry due to their low optical depths may have to be taken into account at slant viewing geometry.
\item These minor condensates may include ones we could reasonably predict the location and distribution of (such as MnS) with an accurate pressure-temperature profile and assumed chemical mixing ratios, and those that we may remain ignorant of, such as photochemically produced hazes, that may have negligible \emph{normal} optical depths.
\end{itemize} 
We assert that transmission spectroscopy will continue to yield abundances of expected chemical species far below those predicted for a `clear' atmosphere, for \hd, and for other planets that may be studied in the future.

\section*{Acknowledgments}

I thank K. Lodders for numerous helpful discussions and for sharing condensation data prior to publication.  I thank M.~S. Marley for comments on the manuscript and also D. Charbonneau, T.~M. Brown, J.~W. Barnes, C.~S. Cooper, and W.~B. Hubbard for interesting discussions.  J.~J.~F. is supported by an NRC Research Associateship.
                                                             

\begin{table}
\centering
\begin{minipage}{80mm}
\caption{Quantities of Intererest for Various Planets}
\begin{tabular}{cccccc}
\hline 
\hline
Planet & $a$ (km) & $H$ (km) & $\eta$ & $H_{cond}$ (km) & $\eta_{cond}$\footnote{Value for $H_{cond}$ and $\eta_{cond}$ (condensate) are computed assuming $H_{cond}=1/3$ $H_{gas}$.}\\
\hline
Earth  &  6400 & 7 & 76 & 2.3 & 132\\
Jupiter& 70000 & 27 & 128& 9 & 221\\
\hd    & 100000 & 440& 38& 147 & 66\\
\T     & 75000 & 185 & 50& 62 & 87\\
\hline
\end{tabular}
\label{table:param}
\end{minipage}
\end{table}

\begin{table}
\centering
\begin{minipage}{80mm}
\caption{Optical depths (at a wavelength of 1 $\mu$m) of lesser condensates in \hd, for an atmosphere of solar composition}
\begin{tabular}{cccccc}
\hline 
\hline
Condensate & Molecular & Limiting & $f$ & normal & Slant\\
 & Weight & Species & & $\tau$ & $\tau$\\
\hline
MgSiO$_3$ & 100.4 & Si/Mg & 7 $\times 10^{-5}$ & 0.5 & 33\\
Na$_2$S & 78.1 & Na & 2 $\times 10^{-6}$ & 0.011 & 0.73\\
NH$_4$H$_2$PO$_4$ & 113 & P & 5.8 $\times 10^{-7}$ & 0.011 & 0.73\\
MnS & 87 & Mn & 6.3 $\times 10^{-7}$ & 0.0093 & 0.61\\
Cr & 85 & Cr & 8.8 $\times 10^{-7}$ & 0.0077 & 0.51\\
KCl & 74.6 & K  & 2.5 $\times 10^{-7}$ & 0.0028 & 0.18\\
ZnS & 97.5 & Zn & 8.5 $\times 10^{-8}$ & 0.0014 & 0.09\\
\hline
\end{tabular}
\label{table:abund}
\end{minipage}
\end{table}

%

\bsp

\label{lastpage}

\end{document}